\documentclass[aps,prl,twocolumn,showpacs,preprintnumbers,amsmath,amssymb]{revtex4}
\usepackage{graphicx}
\usepackage{bm}
\usepackage{dcolumn}

\def\nat#1#2#3{Nature {\bf #1}, #2 (#3)}
\def\prl#1#2#3{Phys. Rev. Lett. {\bf #1}, #2 (#3)}

\def\epjd#1#2#3{Eur. Phys. J. D {\bf #1}, #2 (#3)}
\def\pra#1#2#3{Phys. Rev. A {\bf #1}, #2 (#3)}

\def\epl#1#2#3{Europhys. Lett. {\bf #1}, #2 (#3)}

\def\jpb#1#2#3{J. Phys. B: At. Mol. Opt. Phys. {\bf #1}, #2 (#3)}
\def\physa#1#2#3{Physica A {\bf #1}, #2 (#3)}
\def\pre#1#2#3{Phys. Rev. E {\bf #1}, #2 (#3)}
\def\prb#1#2#3{Phys. Rev.  B {\bf #1}, #2 (#3)}
\def\rmp#1#2#3{Rev. Mod. Phys. {\bf #1}, #2 (#3)}

\def\jpamt#1#2#3{J. Phys. A: Math. Theor. {\bf #1}, #2 (#3)}

\def\noi{\noindent}
\def\bc{\begin{center}}
\def\ec{\end{center}}
 \newcommand{\bea}{\begin{equation}}
 \newcommand{\eea}{\end{equation}\noi}
 \newcommand{\ber}{\begin{eqnarray}}
 \newcommand{\eer}{\end{eqnarray}\noi}
\begin{document}
\title{Critical Casimir force in the superfluid phase: Effect of fluctuations}
\author{Shyamal Biswas$^1$}\email{sbiswas.phys.cu@gmail.com}
\author{J. K. Bhattacharjee$^{2}$}
\author{Himadri S. Samanta$^3$}
\author{Saugata Bhattacharyya$^4$}
\author{Bambi Hu$^{5,6}$}
\vskip0.4cm \affiliation{$^1$Department of Physics, University of Calcutta, 92 Acharya Prafulla Chandra Road, Kolkata-700 009, India
 \vskip0.1cm
$^2$SN Bose National Centre for Basic Sciences, Sector 3, JD Block, Salt Lake, Kolkata-700 098, India
\vskip0.1cm
$^3$Department of Physics and Astronomy, University of Sheffield, Hounsfield Road, Sheffield S3 7RH, UK
\vskip0.1cm $^4$Department of Physics, Vidyasagar College, 39 Sankar Ghosh Lane, Kolkata-700 006, India
\vskip0.1cm $^5$Centre for Nonlinear Studies, and BHKS Joint Centre for Nonlinear and Complex Systems, Hong Kong Baptist University, Kowloon Tong, HK
\vskip0.1cm $^6$Department of Physics, University of Houston, Houston, Texas 77204, USA}

\date{\today}

\begin{abstract}
We have considered the critical Casimir force on a $^4$He film below and above the bulk $\lambda$ point. We have explored the role of fluctuations around the mean field theory in a perturbative manner, and have substantially improved the mean field result of Zandi et al [Phys. Rev. E {\bf 76}, 030601(R) (2007)]. The Casimir scaling function obtained by us approaches a universal constant ($-\frac{\zeta(3)}{8\pi}$) for $T\lesssim 2.13~\text{K}$.
\end{abstract}

\maketitle

\subsection{1. Introduction}
Recently, Garcia $\&$ Chan and Ganshin et al measured the Casimir force induced thinning of the liquid $^4$He film near the bulk $\lambda$ point ($T_\lambda=2.1768~\text{K}$) \cite{chan, chan2}. They obtained a universal scaling function ($\vartheta$) for the critical Casimir force below and above the $\lambda$ point, and observed a dip minimum and a non vanishing constant tail in the $\vartheta$ below the $\lambda$ point. This experiment challenges our understanding of the finite size effects on the films near their bulk critical points. On this issue, the Casimir effects on different critical films have been the subject of a number of experimental \cite{mukhopadhyay,chan,chan-he3,fukuto,chan2,rafai,hertlein} and theoretical \cite{dietrich-he3,kardar,hucht,vasilyev,maciolek,diehl,hasenbusch,dohm,biswas3} works within the last few years.

Although the scaling function was appreciably obtained by the Monte Carlo simulations of Hucht \cite{hucht} and Vasilyev et al \cite{vasilyev} yet this problem is still unsolved analytically. That the confinement of the critical fluctuations may give rise to a (classical) Casimir force was first proposed by Nightingale and Indekeu \cite{nightingale}. Thereafter, a renormalization group calculation for the $\vartheta$ above the $T_\lambda$ was presented by Krech and Dietrich \cite{krech}. For $T<T_\lambda$, a mean field theory with the Ginzburg-Landau (G-L) model was recently presented by Zandi et al \cite{kardar}. They obtained an analytic expression for the $\vartheta$ in terms of the maximum of the superfluid order parameter. By proposing that their mean field calculation could be improved by the confinement of the critical fluctuations (at the Gaussian level), they nicely improved their result only at the $\lambda$ point.

We analytically improve the mean field result of Zandi et al as proposed by them for $T<T_\lambda$ \cite{kardar}. The improvement for $T>T_\lambda$, was already done by Krech and Dietrich even beyond the Gaussian level \cite{krech}. However, we present a physically motivated regularization technique for obtaining the critical Casimir force above the $\lambda$ point. Thus we build a unified picture for the theory of critical Casimir force acting on a $^4$He film below and above the $\lambda$ point. Our theory interestingly predicts the non vanishing constant tail of $\vartheta$ as $-0.0478$, which agrees well with the numerical result of Hucht \cite{hucht} but differs by a factor of five from the experimental value ($-0.24$) \cite{chan2}. Nonetheless, it is a considerable improvement over the mean field calculation which predicts it to be zero \cite{kardar}.

We start from the G-L model. For $T>T_\lambda$, we obtain the free energy in terms of the discrete Fourier modes. The Casimir force is then obtained in the Fisher-de Gennes' form \cite{fisher} by applying the Poisson summation formula \cite{biswas}. Use of this summation formula distinguishes our approach from that of Krech and Dietrich \cite{krech}. For $T<T_\lambda$, we transform the critical fields by introducing the superfluid order parameter, and express the G-L free energy in a decoupled form of the mean field and fluctuating parts. The fluctuating part is treated like that we do for $T>T_\lambda$, and the mean field part is treated in the manner of Zandi et al \cite{kardar}. It is necessary to know the maximum of the order parameter for plotting the mean field part of the Casimir force. Although the graphical solutions of the maximum of the order parameter are exact yet the solutions do not appear in a closed form. We predict a closed form of the maximum of the order parameter from asymptotic analyses, and obtain an approximate mean field Casimir force which matches very well with the exact mean field result \cite{kardar}. Finally we improve the mean field result by adding the contribution of the fluctuating part.

\subsection{2. Free energy of the critical fluctuations for $T>T_\lambda$}
According to the experimental setup $^4$He vapor comes in contact of a plate, and upon liquefaction it forms a film of thickness $238-340 \text{\AA}$ \cite{chan, chan2}. We consider the plate to be along the $x-y$ plane of the co-ordinate system, the area of the film to be $A$, and the thickness of the film to be $L$ along the $z$ direction. Near the $\lambda$ point $^4$He behaves critically, and its local free energy can be written in the G-L framework as
 \bea
F_l=\int d^3{\bf r}\bigg[\frac{1}{2}|\nabla\phi({\bf r})|^2+\frac{a}{2}\big|\phi({\bf r})\big|^2+\frac{b}{4}\big|\phi({\bf r})\big|^4\bigg]
 \eea
where $\phi({\bf r})=\phi_1({\bf r})+i\phi_2({\bf r})$ is a complex scalar critical field at the position vector ${\bf r}=x\hat{i}+y\hat{j}+z\hat{k}$, $a$ is the inverse square of the correlation length ($\xi=\xi_0t^{-\nu}$), $t=T/T_\lambda-1$ is the reduced temperature, $\nu$ is the correlation length exponent, and $b$ is a positive coupling constant \cite{jkb, barber}. The quartic term in Eqn.(1) is neglected in the Gaussian approximation.

Let us first calculate the Casimir force for $T>T_\lambda$. In conformity with the Dirichlet boundary conditions, the Fourier expansion of the critical fields are given by $\phi_{1,2}({\bf r})=\sqrt{\frac{2}{L}}\sum_{n=1}^{\infty}\int\phi_{1,2,n}({\bf
k})\text{sin}\big(\frac{n\pi z}{L}\big)e^{i {{\bf k}.(x\hat{i}+y\hat{j})}}\frac{d^2{\bf k}}{(2\pi)^2}$. In the basis of the Fourier modes, we obtain the partition function ($Z=\int D[\phi_1] D[\phi_2]e^{-F_l/k_BT}$) within the Gaussian approximation, and get the standard form of the free energy ($-k_BT\text{ln}Z$) of the critical fluctuations of the film as \cite{krech}
 \bea
F=2\times\frac{k_BTA}{2}\sum_{n=1}^\infty\int_0^\infty\text{ln}\bigg(k^2+a+\frac{n^2\pi^2}{L^2}\bigg)\frac{kdk}{2\pi}.
 \eea
The factor 2 in the above equation comes from the fact that $\phi$ has two components.

\subsection{3. Critical Casimir force for $T>T_\lambda$}
From Eqn.(2) we get the force acting on the film as
 \bea
f_L=-\frac{\partial F}{\partial L}=2\frac{\pi^2k_BTA}{L^3}S
 \eea
where $S=\sum_{n=1}^\infty\int_0^\infty\frac{n^2}{k^2+a+\frac{n^2\pi^2}{L^2}}\frac{kdk}{2\pi}$. This expression can be recast as
\begin{eqnarray}S&=&\sum_{n=1}^\infty\int_0^\infty\int_0^{\infty}n^2e^{-(k^2+a+\frac{n^2\pi^2}{L^2})t_1}dt_1\frac{kdk}{2\pi}\nonumber\\&=&-\frac{1}{4\pi}\int_0^{\infty}t_1^{-1}e^{-at_1}\frac{\partial}{\partial\tau_1}\sum_{n=1}^{\infty}e^{-n^2\tau_1}dt_1                         \end{eqnarray}
where $\tau_1=\frac{t_1\pi^2}{L^2}$. Using the Poisson summation formula in Eqn.(4), we recast Eqn.(3) as
\begin{eqnarray}
f_L=2&\times&\frac{Ak_BT\pi}{4L^3}\int_0^{\infty}dt_1\frac{e^{-at_1}}{t_1}\bigg(\frac{\sqrt{\pi}}{4\tau_1^{3/2}}
+\frac{\sqrt{\pi}}{2\tau_1^{3/2}}\nonumber\\&\times&\sum_{n=1}^{\infty}e^{-\frac{n^2\pi^2}{\tau_1}}
-\sqrt{\frac{\pi}{\tau_1}}\sum_{n=1}^{\infty}\frac{n^2\pi^2}{\tau_1^2}e^{-\frac{n^2\pi^2}{\tau_1}}\bigg).
\end{eqnarray}
As $L\rightarrow\infty$, only the first term of the parentheses of Eqn.(5) survives. This is the bulk force acting on the film. By the standard analytic continuation technique we get the expression of this bulk force as $f_{\infty}=2\times\frac{Ak_BT}{16}(\frac{a}{\pi})^{3/2}\Gamma(-3/2)$. Subtracting this bulk part from $f_L$ we get the Casimir force in the Fisher-de Gennes form \cite{fisher} $f_C[L,t]=\frac{Ak_BT_\lambda}{L^3}\vartheta(t)$ where $\vartheta(t)$ is the Casimir scaling function which can be expressed in terms of a scaled temperature $\tau=L^{1/\nu}t$ as
\begin{eqnarray}
\vartheta(\tau)=-2\times\frac{1}{8\pi}\sum_{n=1}^{\infty}\bigg(\frac{1}{n^3}+\frac{2\tau^{\nu}}{\xi_0 n^2}+\frac{2\tau^{2\nu}}{n\xi_0^2}\bigg)e^{-\frac{2n\tau^{\nu}}{\xi_0}}.
\end{eqnarray}
We have $\nu=1/2$ in the Gaussian as well as in the mean field approximations \cite{barber}. However, if we want to include the effect of the $|\phi|^4$ term within the above prescription, we must put $\nu=0.67016~(\approx2/3)$ in Eqn.(6) \cite{chan,chan2,barber}. From Eqn.(6) we get the value of the $\vartheta(\tau)$ at the $\lambda$ point as $-\frac{\zeta(3)}{4\pi}=-0.0956$ which matches well with the experimental data obtained by Garcia and Chan \cite{chan}. The same number at the $\lambda$ point was also obtained in Refs.\cite{krech, kardar} with different regularization techniques. We need to know the value of $\xi_0$ for plotting the $\vartheta(\tau)$ against $\tau$. The experimental value of $\xi_0$ for $T>T_\lambda$, varies from $1.2$ to $1.43\text{\AA}$ \cite{ihas,avenel,gasparini}. With no a priory reason we take $\xi_0=1.3\text{\AA}$ for $T>T_\lambda$ \cite{avenel}.

\subsection{4. Free energy of the critical fluctuations for $T<T_\lambda$}
In addressing the situation below the $\lambda$ point we note that the Casimir scaling function looks qualitatively similar to the ultrasonic attenuation and finite size specific heat (i.e. both have a peak below $T_\lambda$) \cite{ferrell, rhee}. We anticipate that the Casimir effect for $T<T_\lambda$ can be thought of as coming from mean field and fluctuating parts. Splitting the ultrasonic attenuation into a sum of mean field and fluctuating parts was the original contribution of Landau and Khalatnikov, and it gives a good account of the ultrasonic attenuation below the $\lambda$ point \cite{landau}. Here we show how a similar approach for the critical Casimir effect can be adopted below the $\lambda$ point.

We return to Eqn.(1) and note that for $T<T_\lambda$, $a$ becomes negative, and accordingly we write $a=-|a|$. This leads to a broken-symmetry, and we handle it by transforming the fields $\phi_2$ and $\phi_1$ to $\psi_2=\phi_2$ and $\psi_1=\phi_1-m(z)$ where $m(z)$ is the superfluid order parameter. The expectation value $<\phi_1>$ is now $z$-dependent because we are considering a finite size system in the $z$-direction and consequently, we expect an inhomogeneity in the superfluid density ($\sim m^2(z)$). The fields $\psi_1,\psi_2$ are such, that $<\psi_i>=0$ ($i=1,2$), and the local free energy in Eqn.(1) in terms of $\psi_1,\psi_2$ becomes
\begin{eqnarray}
F_l&=&\int d^3{(\bf
r)}\bigg[\big[\frac{1}{2}\big(\frac{dm}{dz}\big)^2-\frac{|a|m^2}{2}+\frac{bm^4}{4}\big]+\big[-\frac{d^2m}{dz^2}\nonumber\\&-&|a|m+bm^3\big]\psi_1 +\frac{1}{2}\big[(3bm^2-|a|)\psi_1^2+(\nabla\psi_1)^2\big]\nonumber\\&+&\frac{1}{2}\big[(bm^2-|a|)\psi_2^2+(\nabla\psi_2)^2\big]
+\big[bm\psi_1(\psi_1^2+\psi_2^2)\nonumber\\&+&\frac{b}{4}(\psi_1^2+\psi_2^2)^2\big]\bigg].
\end{eqnarray}
The free energy can be minimized from the condition that $<\frac{\delta F_l}{\delta\psi_1}>=0$, and can be recast from Eqn.(7) as $[-\frac{d^2m}{dz^2}-|a|m+bm^3]+bm[<\psi_2^2>+3<\psi_1^2>]=0$ which can only be solved analytically if we disregard the second square bracketed term by considering the necessary condition that the mean field part dominates over the fluctuating part (i.e. $m^2\gg<\psi_i^2>$). With this consideration we can write an approximate equation for the profile of $m(z)$ as
 \bea
  -\frac{d^2m}{dz^2}-|a|m+bm^3=0.
 \eea
It is to be noted that Eqn.(8) does not minimize the local free energy in Eqn.(7). Hence, the quadratic terms in $\psi_1$ and $\psi_2$ may not be positive. However, Eqn.(8) would minimize the local free energy if we replace $m(z)$ in the quadratic and higher order terms by its bulk value ($\sqrt{|a|/b}$). With all the above considerations (and with Eqn.(8)) the fluctuating and mean field parts of the local free energy become decoupled, and consequently, we recast Eqn.(7) as
 \bea
F_l=F_{mf}+F_o+F_{int}
 \eea
where $F_{mf}=\int d^3{(\bf r)}[\frac{1}{2}(\frac{dm}{dz})^2-\frac{|a|m^2}{2}+\frac{bm^4}{4}]$ is the mean field part, $F_o=\int d^3({\bf r})[\frac{1}{2}2|a|\psi_1^2+\frac{1}{2}(\nabla\psi_1)^2+\frac{1}{2}(\nabla\psi_2)^2]$ is the (Gaussian) fluctuating part, and $F_{int}=\int d^3({\bf r})[\sqrt{|a|b}\psi_1(\psi_1^2+\psi_2^2)+\frac{b}{4}(\psi_1^2+\psi_2^2)^2]$ is the interaction part. We now check that all the quadratic terms in $F_l$ are positive. Hence, Eqn.(8) minimizes the local free energy, and consequently, the quadratic terms in Eqn.(9) on an average dominate over the higher order terms as because $\frac{|a|}{b}\approx m^2\gg<\psi_i^2>$. Evaluation of the partition function from the extremized local free energy in Eqn.(9) leads to
\begin{eqnarray}
Z&=&e^{-\big[\frac{F_{mf}}{k_BT}\big]}Z_o\bigg[1-\big<\frac{F_{int}}{k_BT}\big>_\text{o}+\frac{1}{2}\big<\big(\frac{F_{int}}{k_BT}\big)^2\big>_\text{o}-...\bigg]\nonumber\\&\approx&e^{-\big[\frac{F_{mf}}{k_BT}\big]}Z_oe^{-\big[\big<\frac{F_{int}}{k_BT}\big>_\text{o}-\frac{1}{2}\big<(\frac{F_{int}}{k_BT})^2\big>_\text{o}\big]}
\end{eqnarray}
where $Z_o=\int D[\psi_1]D[\psi_2]e^{-\frac{F_o}{k_BT}}$ is the partition function for the (Gaussian) fluctuating part, and the expectation value $<...>_\text{o}$ is taken with respect to $F_o$. The free energy obtained from Eqn.(10) is given by
\begin{eqnarray}
F=F_{mf}+F_{cf}&+&k_BT\bigg[\big<\frac{F_{int}}{k_BT}\big>_\text{o}\nonumber\\&-&\frac{1}{2}\big<\big(\frac{F_{int}}{k_BT}\big)^2\big>_\text{o}+..\bigg]
\end{eqnarray}
where $F_{cf}=-k_BT\text{ln}Z_o$ is the (Gaussian) fluctuating part of the free energy for $T<T_\lambda$.

\subsection{5. Critical Casimir force for $T<T_\lambda$}
\subsubsection{5.A. Fluctuating contribution}
The fluctuating part of the free energy in Eqn.(11) can be recast in a special form of Eqn.(2) as $F_{cf}=\frac{Ak_BT}{2}\sum_{n=1}^\infty\int_0^\infty\big[\text{ln}[k^2+2|a|+\frac{n^2\pi^2}{L^2}]+\text{ln}[k^2+\frac{n^2\pi^2}{L^2}]\big]\frac{kdk}{2\pi}$ which gives the Casimir scaling function
\begin{eqnarray}
\vartheta_{cf}(\tau)=&-&\frac{1}{8\pi}\bigg[\sum_{n=1}^{\infty}\bigg(\frac{1}{n^3}+\frac{2^{3/2}|\tau|^{\nu}}{\xi_0
n^2}+\frac{4|\tau|^{2\nu}}{n\xi_0^2}\bigg)e^{-\frac{2^{\frac{3}{2}}n|\tau|^{\nu}}{\xi_0}}\bigg]\nonumber\\&-&\frac{\zeta(3)}{8\pi}
\end{eqnarray}
by following the steps from Eqn.(2) to Eqn.(6).

Since $\frac{|a|}{b}\approx m^2\gg<\psi_i^2>$ we can easily check from Eqn.(11) that $F_{mf}\gg F_{cf}\gg <F_{int}>_\text{o}$. Thus we can ignore the $<F_{int}>_\text{o}$ terms in Eqn.(11), and can expect that the Casimir force obtained from the fluctuating part would be much smaller than that obtained from the mean field part.

\subsubsection{5.B. Mean field contribution}
\begin{figure}
\includegraphics{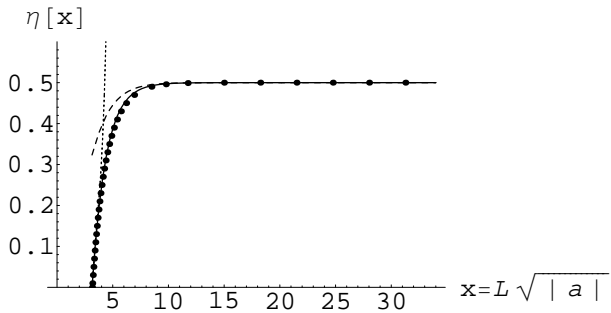}
\caption {The dots are the graphical solutions of $\eta$ in Eqn.(14). The dotted and dashed lines are the asymptotic solutions near $L\sqrt{|a|}\rightarrow\pi$ and $L\sqrt{|a|}\rightarrow\infty$ respectively. The continuous line follow Eqn.(15).}
\label{fig:}
\end{figure}

Let us now evaluate the Casimir scaling function from the mean field part ($F_{mf}$) in Eqn.(11). From the consideration that the order parameter $m(z)$ is smooth and obey the Dirichlet boundary conditions $m(0)=m(L)=0$, $m(z)$ must be symmetric about $z=\frac{L}{2}$ and there would be a single maximum of $m(z)$ at $z=\frac{L}{2}$ for the lowest possible value of the mean field free energy. An analytical expression of the Casimir force ($-\big[\frac{\partial F_{mf}}{\partial L}-\frac{\partial F_{mf}}{\partial L}\big|_{L\rightarrow\infty}\big]$) from the mean field part $F_{mf}$ (and from Eqn.(8)) was nicely obtained in Ref.\cite{kardar} in terms of $\eta=\frac{b}{2|a|}\big(m(\frac{L}{2})\big)^2$ as
\begin{eqnarray}
f_{mf}=-\frac{A|a|^2}{b}\bigg[\frac{1}{4}-\eta(1-\eta)\bigg].
\end{eqnarray}
In Eqn.(13) $\eta$ as well as the maximum of the order parameter is restricted by \cite{kardar,biswas2}
\begin{eqnarray}
L\sqrt{|a|}=\frac{2K\big(\frac{\eta}{1-\eta}\big)}{\sqrt{1-\eta}}
\end{eqnarray}
where $K(x)=\frac{\pi}{2}[1+\frac{x}{4}+\frac{9}{64}x^2+\frac{25}{256}x^3+...]$ is the complete elliptic integral of the first kind. Eqn.(14) gives the allowed range $0\le\eta<\frac{1}{2}$ for the corresponding domain $\pi\le L\sqrt{|a|}<\infty$, and it can be exactly solved by the graphical method \cite{biswas2}. Although the graphical method does not provide $\eta$ in a closed form of $L\sqrt{|a|}$ yet we can do so by the asymptotic analyses near $\eta\rightarrow 0$ and $\eta\rightarrow\frac{1}{2}$. For $L\sqrt{|a|}\rightarrow\infty$, the asymptotic solution of $\eta$ in Eqn.(14) is $\eta\rightarrow\frac{1}{2}\text{tanh}^2\big(\frac{L\sqrt{|a|/2}}{2}\big)$ \cite{biswas2}. On the other hand, for $L\sqrt{|a|}\rightarrow\pi$, the asymptotic solution (up to the third order in $\eta$ in Eqn.(14)) is $\eta\rightarrow\frac{2}{3}\big(\frac{L\sqrt{\bar{a}}}{\pi}\big)^2\big(1-\frac{25}{24}\big(\frac{L\sqrt{\bar{a}}}{\pi}\big)^2+1.04514\big(\frac{L\sqrt{\bar{a}}}{\pi}\big)^4+...\big)$ where $\bar{a}=|a|-\frac{\pi^2}{L^2}$ \cite{biswas2}. Corresponding to the above asymptotic solutions we can take a fitting function for the domain $\pi\le L\sqrt{|a|}<\infty$ as \cite{biswas2, biswas3}
\begin{eqnarray}
\eta(L\sqrt{|a|})=\frac{1}{2}\text{tanh}^2\bigg(\frac{\sqrt{(L^2|a|-\pi^2)/2}}{2}\bigg).
\end{eqnarray}
We see in FIG. 1 that all the asymptotic and graphical solutions match very well with Eqn.(15). Hence, we consider Eqn.(15) as an approximate solution for the rest of this paper.

With the consideration of Eqn.(15), and that $\eta=0$ \cite{kardar} for $0\le L\sqrt{|a|}\le\pi$, one can recast Eqn.(13) in terms of the reduced temperature and the mean field correlation length as
\bea
f_{mf}=\bigg\{\begin{matrix}&-\frac{A|t|^2}{4b\xi_0^4} \ \ \ \ \ \ \ \ \ \ \ \ \ \ \ \ \ \ \ \ \ \ \  \text{for} \ \ \pi\ge L/\xi\ge 0\\
&-\frac{A|t|^2}{4b\xi_0^4}\text{sech}^4\bigg(\sqrt{\frac{\big(\frac{L}{\xi}\big)^2-\pi^2}{8}}\bigg) \ \ \ \text{for} \ \ L/\xi\ge\pi.
\end{matrix}
\eea
From Eqn.(16) a dip minimum with discontinuous slope is expected to occur at $L\sqrt{|a|}=\pi$. This point is fitted to the experimental dip at $\tau=-9.7$ \cite{chan2}. The modifications to Eqn.(16) would come from the higher order fluctuating terms, and the primary correction would be to keep the form of the $f_{mf}$ unaltered with the mean field $\xi$ be replaced by $\xi=\xi_0|t|^{-\nu}$ where $\nu$ to the lowest order in $b$ is $\frac{1}{2}+\frac{b}{2}(n+2)$ \cite{barber}. Using the fixed point value of $b$ we can get the usual $\nu$ at one loop order. We can safely assume that the effect of the different loops will be to make $\xi=\xi_0 t^{-\nu}$ with $\nu$ acquiring the value $0.67016~(\approx2/3)$ correct to all orders \cite{barber, chan2}. From Eqn.(16) we get the modified mean field Casimir scaling function in terms of $\tau=L^{1/\nu}t$ as
\begin{equation}
\vartheta_{mf}(\tau)=\bigg\{\begin{matrix}-\frac{|\tau|^{2}}{4b\xi_0^4k_BT_\lambda}\ \ \ \ \ \ \ \ \ \ \ \ \ \ \ \ \ \ \ \  \text{for}\ -9.7\le\tau\le0\\-\frac{|\tau|^{2}}{4b\xi_0^4k_BT_\lambda}\text{sech}^4\bigg(\sqrt{\frac{\frac{|\tau|^{2\nu}}{\xi_0^2}-\pi^2}{8}}\bigg)\ \text{for}\ \tau\le-9.7.
\end{matrix}
\end{equation}
If we plot Eqn.(17) we must get almost the same result as obtained by Zandi et al \cite{kardar}. However, we need to improve the mean field result (in Eqn.(17)) by the confinement of the critical fluctuations as proposed by them.

\subsubsection{5.C. Improvement of the mean field result}
The net Casimir scaling function for $T<T_\lambda$ is obtained from Eqns.(12) and (17), and is given by
  \bea
\vartheta(\tau)=\vartheta_{mf}(\tau)+\vartheta_{cf}(\tau).
  \eea
We plot the right hand sides of Eqns.(6) and (18) in FIG. 2. For $T>T_\lambda$, our theory matches very well with the experimental data of Garcia $\&$ Chan \cite{chan}. From FIG. 2 we also see that the $\vartheta$ approaches a constant $-\frac{\zeta(3)}{8\pi}=-0.0478$ for $\tau\lesssim -75.6~\text{\AA}^{1/\nu}$ and (with $L=238\text{\AA}$) for $T\lesssim 2.13~\text{K}$ as well.

Although our theory for $T<T_\lambda$, does not match very well with the experimental data yet it predicts the basic nature of the critical Casimir force which is characterized by a non vanishing constant tail \cite{chan2}. It is of course clear from FIG. 2 that the inclusion of the effect of the critical fluctuations substantially improves the exact mean field result of Zandi et al \cite{kardar}.

\begin{figure}
\includegraphics{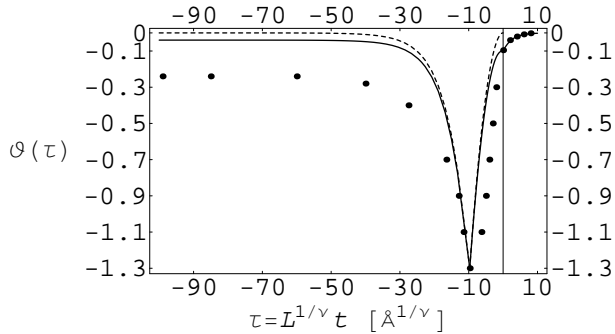}
\caption {The continuous line for $\tau>0$, follows Eqn.(6) with $\xi_0=1.3\text{\AA}$ \cite{avenel}, and that for $\tau<0$, follows Eqn.(18) with $\xi_0=1.4593\text{\AA}$, $\frac{1}{4b\xi_0^4k_BT_\lambda}=0.0133547$ and $\nu=0.67016$. The dashed line represents exact mean field calculation of Zandi et al \cite{kardar} with $\xi_0=1.4593\text{\AA}$, $\frac{1}{4b\xi_0^4k_BT_\lambda}=0.0138166$ and $\nu=0.67016$. A few experimental points are taken from Ref.\cite{chan} for $\tau>0$, and from Ref.\cite{chan2} for $\tau<0$.} \label{fig:}
\end{figure}

\subsection{6. Conclusions}
Complementing the numerical \cite{hucht,vasilyev,hasenbusch} and analytical \cite{kardar,krech,maciolek,diehl,dohm,biswas3} works we have given a unified theory for the the critical Casimir force below and above the lambda point in a single framework. In particular, we have explored the effect of the critical fluctuations in the Gaussian level over the mean field contribution \cite{kardar}.

The tail of the scaling function approaches a non-vanishing constant $-\frac{\zeta(3)}{8\pi}=-0.0478$ owing to the consideration of the two component ($\psi_1, \psi_2$) critical fluctuations. Although this constant is closer to the numerical simulation result obtained by Hucht \cite{hucht} yet it is nearly one fifth of that obtained by the experimentalists \cite{chan2}. On the other hand, this constant is zero in the mean field level \cite{kardar}. Hence, our calculation of the Casimir scaling function goes beyond that of Zandi et al \cite{kardar}, and compares favorably with the numerical simulation of Hucht \cite{hucht} and the experimental data of Garcia $\&$ Chan \cite{chan} and Ganshin et al \cite{chan2}.

It should be mentioned that the theory of the Casimir force for $T<T_\lambda$, was also improved by Zandi, Rudnick and Kardar with the considerations of confinement of the Goldstone modes $\&$ surface fluctuations \cite{kardar2}, and that the thinning of the liquid $^4$He film was first (but admittedly not very precisely) observed by Dionne and Hallock \cite{dionne}.

While the Casimir force for the quantum fluctuations of the electromagnetic field is observed within $10^{-12}\text{~N}$ \cite{mohideen} the critical Casimir force considered by us is observed  within $10^{-3}\text{~N}$ \cite{chan,chan2}. The confinement of the classical (critical) fluctuations of course is much stronger than that of the quantum (vacuum) fluctuations.

The experimental dip of the $\vartheta$ has been adjusted with the value of $b$ which nobody has determined (for the film) so far from the theoretical point of view. How to determine the parameter $b$ for the film and to calculate the Casimir force by considering the coupling between the mean field and fluctuating parts remain to this day as open problems.

\subsection{7. Acknowledgments}
S Biswas acknowledges the hospitalities and the financial supports of the `Department of Theoretical Physics in IACS' and the `CNS in Hong Kong Baptist University' where he did integral parts of the research work of this paper. His final part of the research work has been sponsored by the University Grants Commission [UGC] under the UGC-Dr. D.S. Kothari Postdoctoral Fellowship Scheme [No.F.4-2/2006(BSR)13-280/2008(BSR)].

\end{document}